\documentclass[reprint,prl,showpacs,twocolumn,nobibnotes]{revtex4}
\usepackage{amsmath}
\usepackage{amssymb}
\usepackage{graphicx}
\usepackage{color}

\newcommand{\pt}{ \mathcal{PT}}

\newcommand{\be}{\begin{equation}}
\newcommand{\ee}{\end{equation}}

\begin{document}

\title{PT-symmetry breaking and laser-absorber modes in optical scattering systems}

\author{Y.~D.~Chong}

\thanks{These authors contributed equally to this work.}

\author{Li~Ge}

\thanks{These authors contributed equally to this work.}

\author{A.~Douglas~Stone}

\affiliation{Department of Applied Physics, Yale University, New
  Haven, Connecticut 06520}


\begin{abstract}
Using a scattering matrix formalism, we derive the general scattering
properties of optical structures that are symmetric under a combination of
parity and time-reversal ($\pt$).  We demonstrate the existence of a
transition beween $\pt$-symmetric scattering eigenstates, which are
norm-preserving, and symmetry-broken pairs of eigenstates exhibiting net
amplification and loss.  The system proposed by Longhi \cite{Longhi}, which
can act simultaneously as a laser and coherent perfect absorber, occurs at
discrete points in the broken symmetry phase, when a pole and zero of the
S-matrix coincide.
\end{abstract}

\maketitle

Schr\"odinger equations that violate time-reversal symmetry due to a
non-Hermitian potential, but retain combined $\pt$ (parity-time) symmetry,
have been extensively studied since the work of Bender
\textit{et.~al.}~\cite{Bender,Bender2}, who showed that such systems can
exhibit real energy eigenvalues, suggesting a possible generalization of
quantum mechanics.  Moreover, $\pt$-symmetric systems can display a
spontaneous breaking of $\pt$-symmetry, at which the reality of the
eigenvalues is lost \cite{Bender, Makris}.  Although $\pt$-symmetric
quantum mechanics remains speculative as a fundamental theory, the idea has
been fruitfully extended to wave optics \cite{Makris,Musslimani,Guo,Ruter}.
The classical electrodynamics of a medium with loss or gain breaks
$\mathcal{T}$ symmetry in the mathematical sense, although the underlying
quantum electrodynamics is of course $\mathcal{T}$-symmetric.  $\pt$
symmetry is maintained in optical systems by means of balanced gain and
loss regions that transform into one another under parity; thus the
combined $\pt$ operation, which also interchanges loss and gain, leaves the
system invariant.

We show in this Letter that the \textit{scattering} behavior of a
general $\pt$-symmetric system can exhibit one or multiple spontaneous
symmetry-breaking transitions.  This result applies to arbitrarily
complex $\pt$-symmetric scattering geometries, whereas earlier works
on optical $\pt$ symmetry breaking were inherently restricted to
waveguide (paraxial) geometries \cite{Makris,Musslimani,Guo,Ruter} for
which resonances in the propagation direction play no role.  In
addition, we elucidate the properties of certain singular solutions
occurring in such systems, recently studied for special cases by
several authors \cite{Mostafazadeh,Schomerus,Longhi}, where a pole and
a zero of the scattering matrix (S-matrix) coincide at a specific real
frequency.  A real-frequency pole corresponds to the threshold for
laser action, while a real-frequency zero implies the reverse process
to lasing, in which a particular incoming mode is perfectly absorbed.
A device exhibiting the latter phenomenon, which does not require
$\pt$-symmetry, has been termed a ``coherent perfect absorber'' (CPA)
\cite{CPA}.  A $\pt$-symmetric scatterer, at these singular points,
can function simultaneously as a CPA and a laser at threshold, as
noted by Longhi \cite{Longhi}.  The present work establishes the
CPA-laser points as special solutions in a wider ``phase'' of
$\pt$-broken scattering eigenstates.  We identify signatures of both
the $\pt$-breaking transition and the CPA-lasing points, for several
exemplary and experimentally-feasible geometries.

{\it S-matrix properties} -- Consider an optical cavity coupled to a
discrete set of scattering channels, denoted $\mu = 1, 2, \cdots$.
Incoming fields enter via the input channels, interact in the cavity,
and exit via the output channels.  For simplicity, we focus on the
scalar wave equation, which directly describes one-dimensional (1D)
and two-dimensional (2D) systems.  A steady-state scattering solution
for the electric field, $E(\vec{r})$, obeys
\begin{equation}
  \left[\nabla^2 + n^2(\vec{r})\, (\omega^2/c^2)\right] E(\vec{r}) = 0.
  \label{wave equation}
\end{equation}
For amplifying or dissipative media, $n(\vec{r})$ is complex.  The
frequency $\omega$ is real for physical processes but can be usefully
continued to the complex plane.  Henceforth, we set $c = 1$.  Outside
the cavity, $E$ has the form
\begin{equation}
  E(\vec{r}) = \sum_\mu \left[\psi_\mu u_\mu^{\textrm{in}}(\vec{r},\omega) +
    \varphi_\mu u^{\textrm{out}}_\mu(\vec{r},\omega) \right].
  \label{Eoutside}
\end{equation}
Here $u_\mu^{\textrm{in}}(\vec{r},\omega)$ and
$u^{\textrm{out}}_\mu(\vec{r},\omega)$ denote the input and output channel
modes, whose exact form depends on the scattering geometry (e.g.~plane
waves, spherical/cylindrical waves, or waveguide modes).  The complex input
and output amplitudes, $\psi_\mu$ and $\varphi_\mu$, are related by the
S-matrix:
\begin{equation}
  \sum_{\nu} S_{\mu\nu}(\omega)\, \psi_\nu = \varphi_\mu.
  \label{S-matrix}
\end{equation}
For all values of $\omega$, $S(\omega)$ is a (complex) symmetric
matrix \cite{reciprocity}; if $n(\vec{r})$ and $\omega$ are real, it
is also unitary.

We now review the properties of $S(\omega)$ for a complex,
$\pt$-symmetric system \cite{Schomerus}.  (Henceforth, $\mathcal{P}$
will refer to any linear symmetry operation such that $\mathcal{P}^2 =
1$, including not only parity, but also $\pi$ rotations and
inversion.)  First, observe that the $\mathcal{T}$-operator maps
incoming channel modes to outgoing ones:
\begin{equation}
  \mathcal{T} \; u_\mu^{\textrm{in}}(\vec{r},\omega) =
  u_\mu^{\textrm{out}}(\vec{r},\omega^*),
  \label{channel time reversal}
\end{equation}
where $\mathcal{T}$ is the anti-linear complex conjugation operator.  The
generalized parity operator $P$ is a linear operator acting on scalar fields
satisfying
\begin{equation}
  P \, u_\mu^{\textrm{in/out}}(\vec{r},\omega) = \sum_\nu \mathcal{P}_{\nu\mu}
  u_\nu^{\textrm{in/out}}(\vec{r},\omega),
  \label{P matrix}
\end{equation}
where the system-dependent matrix $\mathcal{P}$ mixes the channel
functions, but never transforms between incoming and outgoing
channels.  Note that $(\pt)^2 = 1$.

If the system is $\pt$-symmetric, for any solution (\ref{Eoutside})
there exists a valid solution $(P\mathcal{T}) E(\vec{r})$ at frequency
$\omega^*$:
\begin{equation}
  \sum_\mu \left[ (\pt \varphi)_{\mu} \,
    u_\mu^{\textrm{in}}(\vec{r},\omega^*)\; +\; (\pt \psi)_{\mu} \,
    u^{\textrm{out}}_\mu(\vec{r},\omega^*) \right].
\end{equation}
Comparing this to (\ref{Eoutside}) and (\ref{S-matrix}), we conclude
that
\begin{equation}
  (\pt) \, S(\omega^*) \, (\pt) = S^{-1}(\omega).\quad
  \label{S matrix PT}
\end{equation}
This is the fundamental relation obeyed by $\pt$-symmetric S-matrices,
and we will now show that it has important implications for the
eigenvalue spectrum.

Multiplying both sides of (\ref{S matrix PT}) by an eigenvector
$\psi_n$ of $S(\omega)$ with eigenvalue $s_n$ gives
\begin{equation}
  S(\omega^*) \, \left(\pt\psi_n\right) = \frac{1}{s_n^*} \,
  \left(\pt\psi_n\right). \label{eq:S*eigen}
\end{equation}
Hence, the inverse of the complex conjugate of any eigenvalue of
$S(\omega)$ is an eigenvalue of $S(\omega^*)$.  For real $\omega$,
this implies that $|\det S(\omega)| = 1$, just as for S-matrices
having pure $\mathcal{T}$ symmetry, which are unitary.  Unitarity
imposes a stronger constraint: \textit{each} eigenvalue is unimodular,
so unitary S-matrices do not have poles or zeros for real $\omega$.
Both $\mathcal{T}$ and $\pt$ symmetry imply that poles and zeros occur
in complex conjugate pairs.

{\it Symmetry-breaking transition} -- Eq.~(\ref{S matrix PT}) is a
weaker constraint than unitarity, and can be satisfied in two ways:
either each eigenvalue is itself unimodular, or the eigenvalues form
pairs with reciprocal moduli.  These two possibilities correspond to
symmetric and symmetry-broken scattering behavior.  In 1D, there are
just two scattering eigenvectors, so the entire S-matrix is either in the
symmetric or broken-symmetry ``phase''.  (In higher dimensions, as we
will see, the transition can occur in different eigenspaces of the
S-matrix, so that the system can be in a mixed phase; initially
we focus on 1D.)  Let us denote the eigenvectors by $\psi_\pm$.  In
the symmetric phase, each $\psi_\pm$ is itself $\pt$-symmetric,
i.e.~$\pt\psi_\pm \propto \psi_\pm$, so the eigenstate exhibits no net
amplification nor dissipation ($|s_\pm| = 1$). In the broken-symmetry
phase, $\psi_\pm$ is not itself $\pt$-symmetric but the {\it pair}
satisfies $\pt$ by transforming into each other: $ \pt\psi_\pm =
\psi_{\mp}$, where $s_{\pm} = 1/s_\mp^*$.  Each eigenstate in
the pair spontaneously breaks $\pt$ symmetry; one exhibits
amplification, and the other dissipation.  Similar to transitions in
Hamiltonian systems, such as the transition from real to complex
eigenvalues of $\pt$-symmetric Hamiltonians, the scattering transition
can be induced by tuning the parameters of $S(\omega)$ \cite{Berry}.

\begin{figure}
  \centering\includegraphics[width=6.7cm]{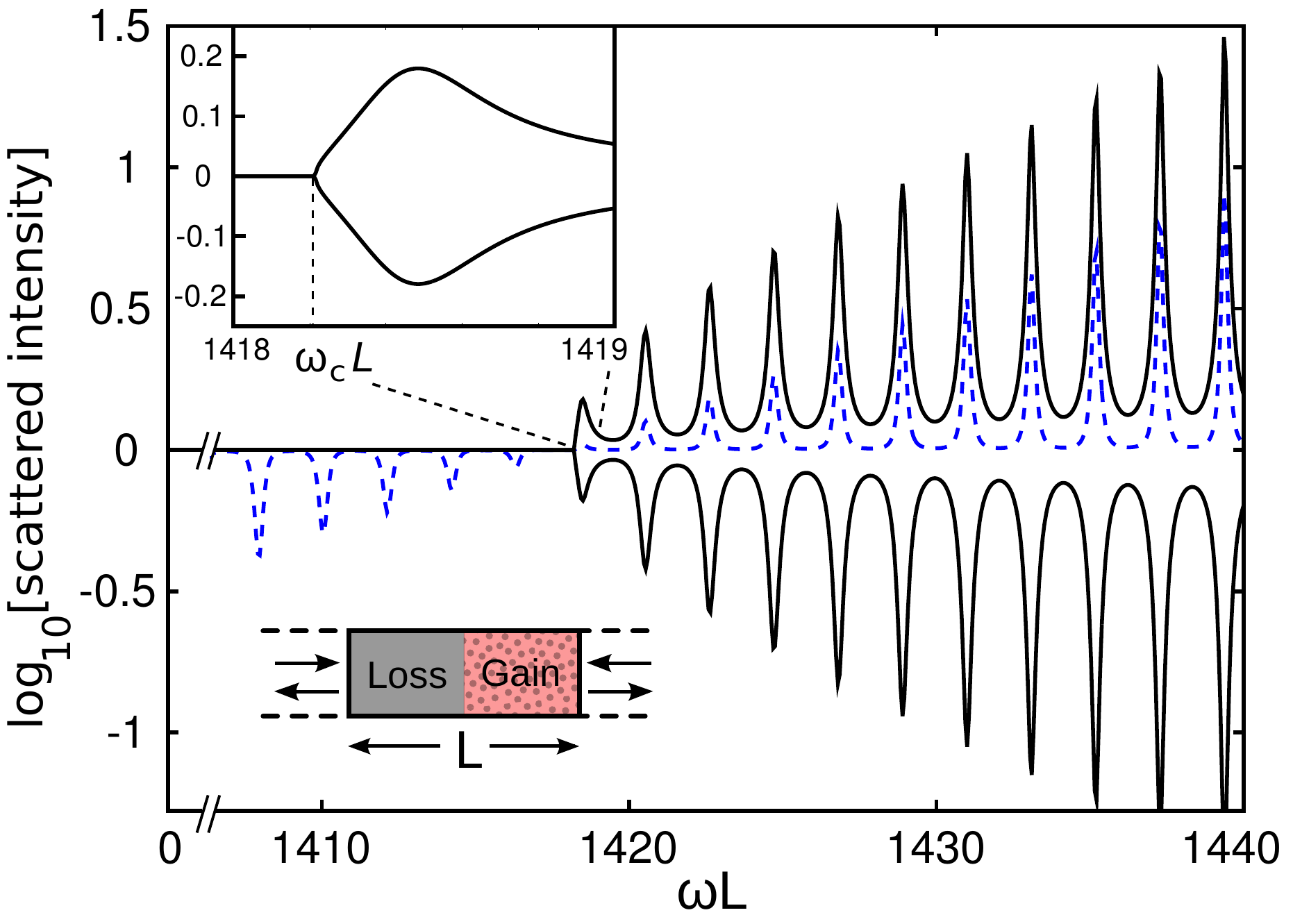}
  \caption{(Color online) Semilog plot of S-matrix eigenvalue
    intensities $\log_{10}|s_\pm|^2$ versus frequency $\omega$ (solid
    curves), for a 1D $\pt$-symmetric system of length $L$ with
    balanced refractive index $n = 3 \pm 0.005i$ in each half (inset).
    The $\pt$ symmetry is spontaneously broken at $\omega_c \approx
    1418.21 / L$.  Dashed curves show the minimum scattered intensity
    for equal-intensity input beams of variable relative phase; in the
    broken-symmetry regime, net amplification always occurs.  }
  \label{fig:onedpt}
\end{figure}

Having shown that a $\pt$-breaking transition can take place in the
scattering behavior of $\pt$-symmetric systems, we turn now to some
concrete examples to see how it occurs.  First, consider an arbitrary
1D system with $\pt$-symmetric $n(x)$; its S-matrix is parameterized
by
\begin{equation}
S = \left(\begin{array}{c c} r_L & t \\ t & r_R
       \end{array}\right)
       = t \left(\begin{array}{c c} (1-|t|^{-2})\frac{1}{ib} & 1 \\ 1 &
         ib
       \end{array}\right),
       \label{eq:Sgeneral}
\end{equation}
where $b \equiv -ir_R/t \in \mathbb{R}$; $r_R,r_L$ are the reflection
amplitudes from right and left; and $t$ is the (direction-independent)
transmission amplitude.  In general $|r_R| \neq |r_L|$, but their relative
phase must be $0$ or $\pi$.  Although $S$ depends on three real numbers,
$\{\rm{Re}(t),\rm{Im}(t),b\}$, its eigenvalues only depend on two, $|t|$
and $b$, for we can scale out the phase of $t$.  One can show that the
criterion for the eigenvalues of $S$ to be unimodular is:
\begin{equation}
  |(r_L-r_R)/t| \, \equiv \, B(\omega,\tau) \, \leq \, 2.
  \label{eq:bifurcation3}
\end{equation}
For fixed $\textrm{Re}\{n(x)\}$, we write $B$ as a function of $\omega$ and
a $\mathcal{T}$-breaking parameter $\tau$.  On varying $\omega$ and/or
$\tau$, violating (\ref{eq:bifurcation3}) brings us into the
broken-symmetry phase.

\begin{figure}
  \centering\includegraphics[width=6.4cm]{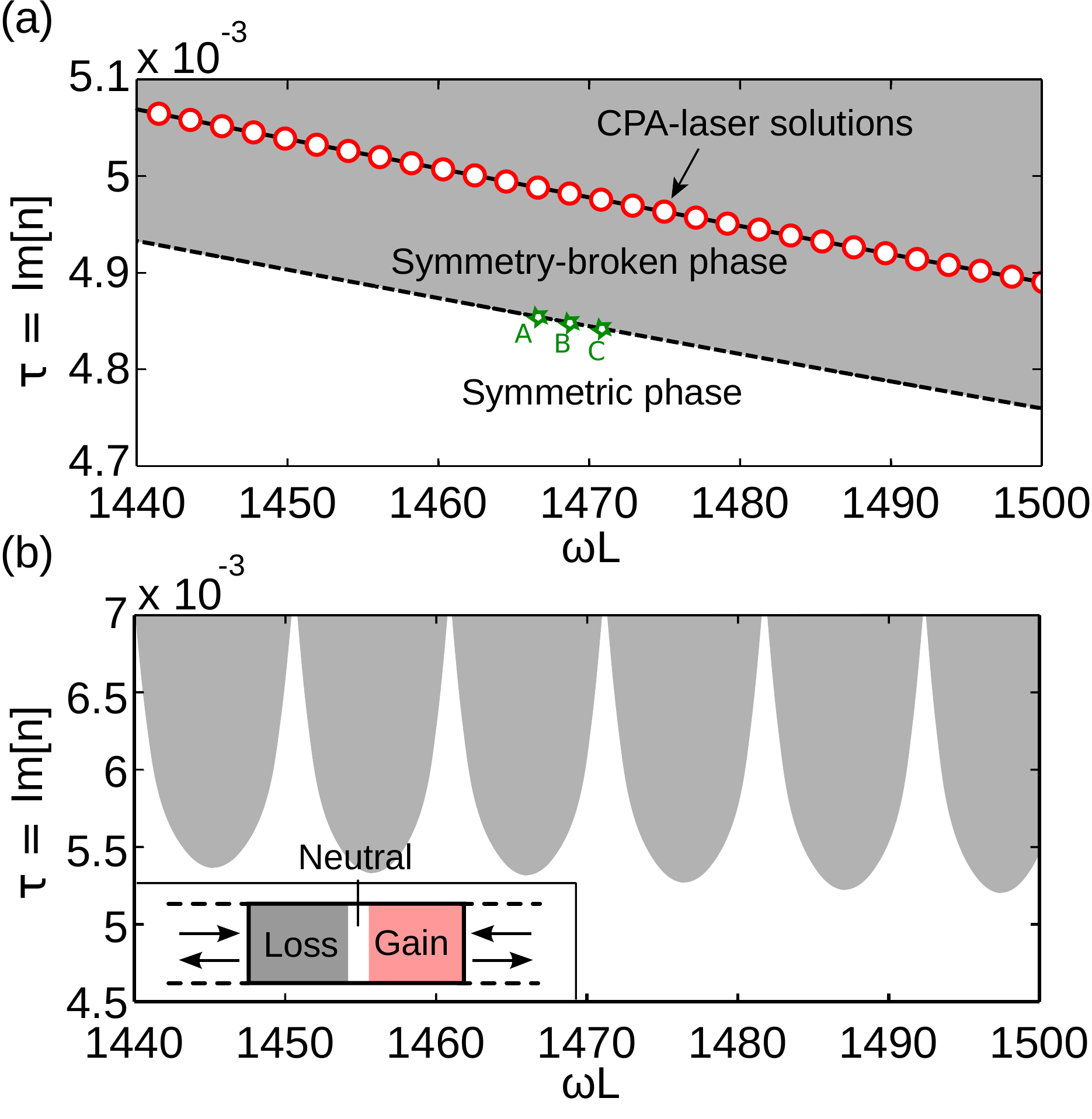}
  \caption{(Color online) Phase diagram of S-matrix eigenvalues for 1D
    $\pt$-symmetric systems.  The structure in (a) is the same as in
    Fig.~\ref{fig:onedpt}.  The structure in (b) has a real-index
    region of length $L/10$ in the center, with all other parameters
    the same as in (a). It shows re-entrant behavior as $\omega$ is varied.
    White areas show the $\pt$-symmetric phase,
    and grey areas show the $\pt$-broken phase; circles in (a) show
    the CPA-laser solutions.  In (a), the dashed line shows the
    approximate phase boundary given by Eq.~(\ref{eq:phaseborder});
    the points A, B, and C on this boundary are referenced in
    Fig.~\ref{fig:ZIflow}.  For fixed $\omega$, stronger
    $\mathcal{T}$-breaking always favors broken symmetry.}
  \label{fig:compare3layers}
\end{figure}

Fig.~\ref{fig:onedpt} shows how the transition occurs by varying
$\omega$, in a simple slab of total length $L$ with fixed $n=n_0 \pm
i\tau$ in each half.  The critical frequency can be shown to be
\begin{equation}
  \omega_c \;\approx\; \ln\!\left(2n_0/\tau\right) \, c \, /\tau L.
  \label{eq:phaseborder}
\end{equation}
The resulting ``phase diagram'' is shown in
Fig.~\ref{fig:compare3layers}(a).  The discrete points in the
broken-symmetry phase correspond to the CPA-laser solutions which we
will soon discuss; these lie along a line given by the equation
\cite{Mostafazadeh}
\begin{equation}
  \omega \approx \omega_c + (c/\tau L)\,\ln[(n_0^2+1)/(n_0^2-1)].
\end{equation}
The strong oscillations of $\log|s_\pm|$ in Fig.~\ref{fig:onedpt}
occur as the system crosses this curve.  Without fine-tuning $n_0$,
the system will not hit one of the CPA-laser solutions exactly; hence,
varying $\omega$ does not bring $\log|s_\pm|$ to $\pm \infty$, though
it can become quite large.

With different $n(x)$, more complicated phase diagrams are possible.
Fig.~\ref{fig:compare3layers}(b) shows the effect of a real-index
region inserted between the gain and loss regions.  As we increase
$\omega$, internal resonances in the real-index region cause the
trapping time to oscillate with $\omega$; hence the {\it effective}
$\mathcal{T}$-breaking perturbation strength oscillates, and the
system re-enters the symmetric phase periodically.

In these 1D structures, the symmetry-breaking transition can be probed
by measuring the total intensity scattered by equal-intensity input
beams of variable relative phase $\phi$.  In the symmetric phase, two
particular values of $\phi(\omega,n)$ correspond to unimodular
S-matrix eigenstates; for other values of $\phi$, both net
amplification and dissipation can be observed in the total output
$|S\psi|^2$, since the system is not unitary.  In the broken-symmetry
phase, where $|s_+| > 1$, it can be shown that
\begin{equation}
  |S\psi|^2 = 1 + |a|^2\left(|s_+|^2+|s_+|^{-2} - 2\right) > 1,
\end{equation}
where $a = \psi_1^T\psi/\psi_1^T\psi_1$.  Thus all values of $\phi$
for balanced inputs $\psi$ give rise to amplification
(Fig.~\ref{fig:onedpt}).  This signature of the $\pt$ transition
should be easily observable.

\begin{figure}
  \centering\includegraphics[width=6.4cm]{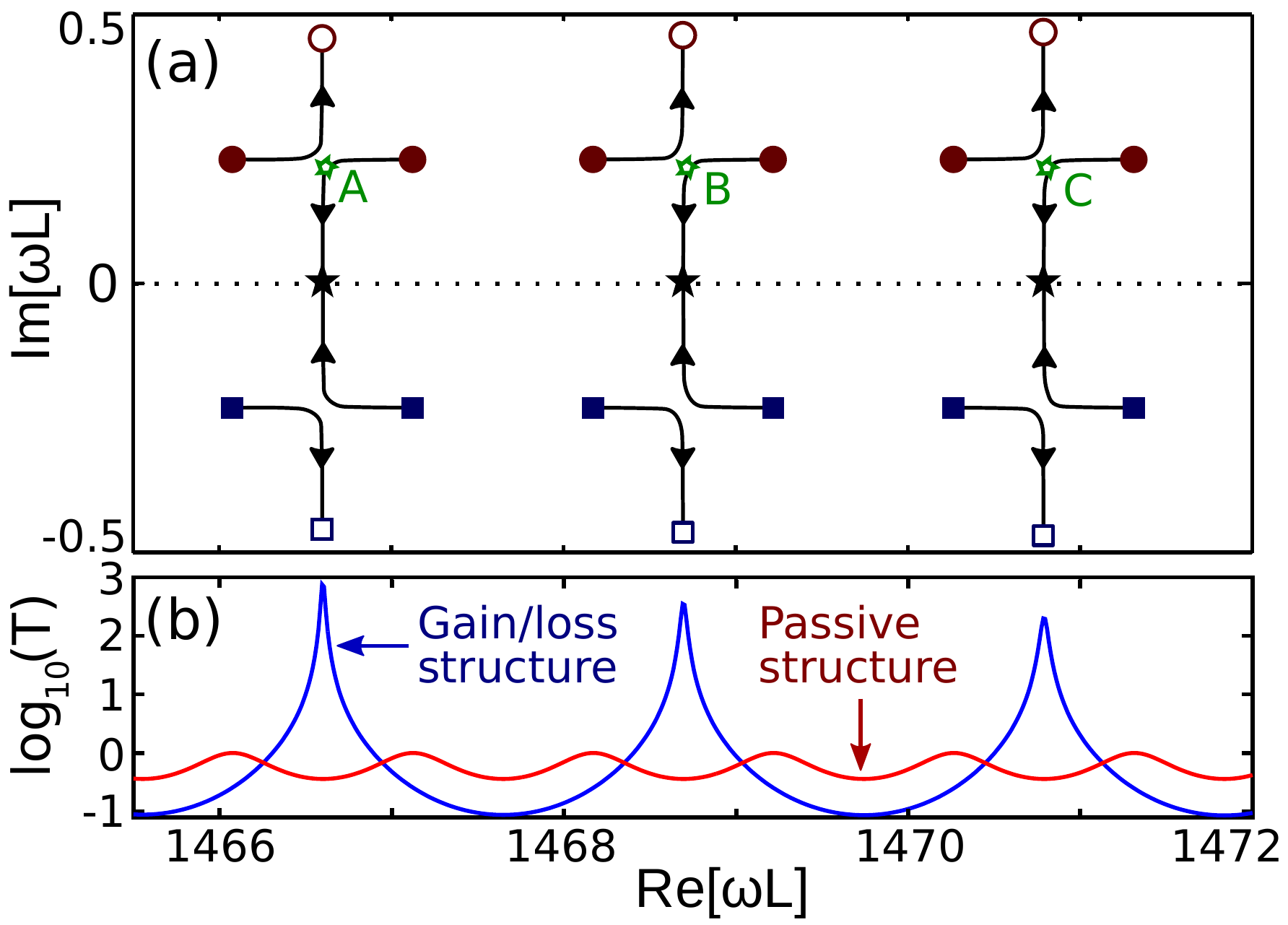}
  \caption{(Color online) (a) Trajectories of S-matrix poles and zeros
    for the two-layer system of Fig.~\ref{fig:onedpt}.  Filled circles
    and squares show the zeros and poles for $\tau = |\text{Im}[n]|
    =0$; solid curves show their trajectories as $\tau$ is increased.
    Filled stars show CPA-laser solutions where zeros and poles meet
    on the real axis.  For the middle set of trajectories, this
    happens at $\omega L = 1468.7$, $\tau=4.982\times10^{-3}$.  The
    transition occurs at the ``anti-crossings'', indicated by the
    representative points A, B, and C, which match the values of
    $\{\textrm{Re}[\omega], \tau\}$ marked in
    Fig.~\ref{fig:compare3layers}. (b) Log of the transmittance for
    $\tau = 0.005$ and $\tau = 0$, showing the doubled free spectral
    range in the CPA-laser. }
  \label{fig:ZIflow}
\end{figure}

Fig.~\ref{fig:ZIflow} shows the behavior of the poles and zeros in the
complex plane for the simple 2-layer structure.  When $\tau = 0$, the
poles and zeros of the $\mathcal{T}$-symmetric S-matrix are located
symmetrically around the real axis, and can be labeled by the parity
of the corresponding eigenvectors. As $\tau$ increases from zero,
$\pt$ symmetry requires that the zeros and poles move symmetrically in
the complex plane.  Neighboring zeros (poles) approach each other
pairwise and undergo an anti-crossing, causing strong parity mixing.
The horizontal motion of the poles and zeros corresponds to the
symmetric phase of the scatterer, and the vertical motion to the
broken-symmetry phase.  The anti-crossing corresponds to the phase
boundary, as can be seen by comparing the points labeled A, B and C in
Figs.~\ref{fig:compare3layers} and \ref{fig:ZIflow}.

{\it CPA-Laser} -- The CPA-laser points occur when a pole and a zero
of the S-matrix coincide on the real axis, as shown in
Fig.~\ref{fig:ZIflow}.  This corresponds to the singular case in which
$|s_n| \to 0$, implying $|1/s^*_n| \to \infty$, while their product
remains unity. Physically, the scattering system behaves
simultaneously as a laser at threshold and a CPA \cite{Longhi}.
Fig.~\ref{fig:ZIflow} also indicates a very interesting property of
these solutions: only half the zero-pole pairs flow upwards and reach
the real axis, while the other half go off to infinity and do not
produce laser-absorber modes; thus, the CPA-laser lines have {\it
  twice} the free spectral range of the passive cavity resonances.
For $\omega L \gg 1$, they occur at frequencies $\omega_m \approx
(2m+0.5)\pi / (n_0L)$, where $m$ is an integer, exactly half-way
between alternate pairs of passive cavity resonances.  This property
of the CPA-laser should be straightforward to demonstrate
experimentally.

In a general $\pt$-symmetric system, the poles and zeros are related
by complex conjugation, so any $\pt$ system that is a laser {\it must}
be a CPA-laser.  Even away from the CPA-laser singularities, the
$\pt$-symmetric cavity in the broken symmetry phase is a unique
interferometric amplifier: for coherent input radiation not
corresponding closely to the damped S-matrix eigenvector,
amplification takes place---in particular, this typically occurs for
one-sided illumination.  However, coherent illumination from both
sides, with appropriate relative phase and amplitude corresponding to
the damped eigenvector, leads instead to strong absorption. The
CPA-laser is the extreme case, where the gain/loss contrast is
infinite.

The possibility of lasing in a $\pt$ system is not obvious, as naively one
might argue that photons traversing the device should experience no net
gain.  The presence of spontaneous symmetry breaking invalidates this
argument.  Photons in amplifying modes spend more time in the gain region
than the loss region; hence, as the gain/loss parameter $\tau$ is
increased, it eventually becomes possible for one mode to overcome the
outcoupling loss and lase.  As a self-organized {\it oscillator}, the
CPA-laser will automatically emit in the amplifying eigenstate.

\begin{figure}
  \centering\includegraphics[width=7cm]{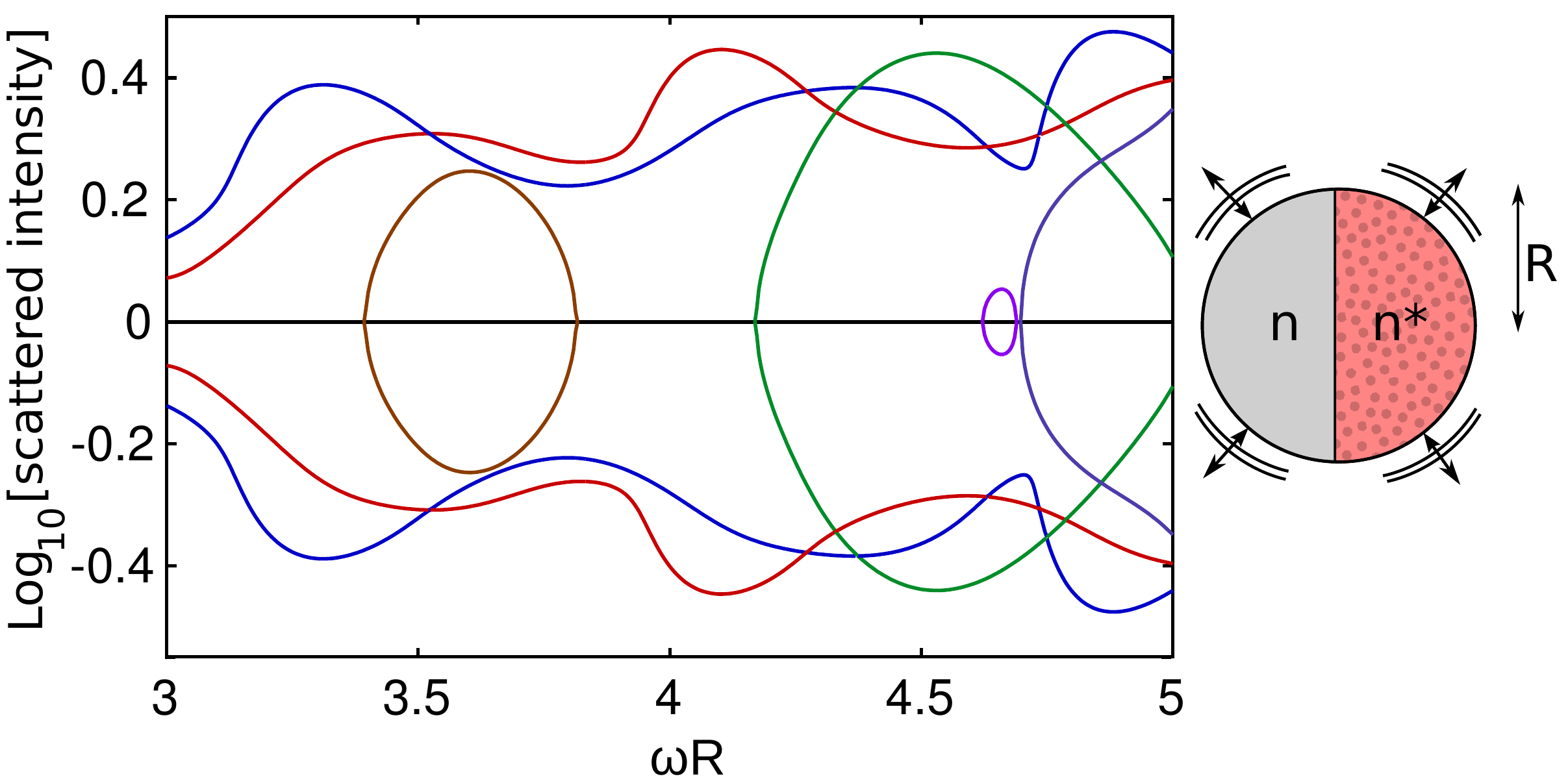}
  \caption{(Color online) Semilog plot of S-matrix eigenvalue
    intensities $\log_{10}|s|^2$ versus frequency $\omega$, for a 2D
    $\pt$-symmetric disk of radius $R$ and refractive indices $1.5 \pm
    0.1i$.  The S-matrix is truncated to angular momenta $|m| \le 20$.
    Most of the 41 eigenvalues are unimodular at these frequencies;
    those with higher average $m$, near resonances of the passive
    disk, show greater symmetry breaking.  The colors distinguish
    several different pairs of $\pt$-broken eigenvalues.}
  \label{fig:twodpt}
\end{figure}

{\it Disk $\pt$-scatterer} -- Consider any 2D $\pt$-symmetric body in
free space.  Scattering in this system is naturally described using
angular momentum channel functions,
\begin{equation}
  u_m^{\textrm{in/out}}(r,\phi,\omega) = H^\mp_m(\omega r)\, e^{\pm i m \phi},
  \; \; m =0,\pm1,\cdots
\end{equation}
For a region of linear dimensions $R$ and average index $\bar{n}$,
only states with $m \lesssim \bar{n} kR$ are significantly scattered,
so we can truncate the infinite S-matrix to $N \gtrsim \bar{n} kR$
channels.  As this S-matrix is tuned, multiple pairs of eigenstates
can undergo $\pt$-breaking, so the phases of the S-matrix are indexed
by the number of pairs of $\pt$-broken eigenstates.  In
Fig.~\ref{fig:twodpt}, we show this behavior for a disk with balanced
semicircular gain and loss regions, with the S-matrix calculated using
the R-matrix method \cite{CPA,wigner}.  The symmetry-breaking is
non-monotonic in $\omega$, similar to the 1D example of
Fig.\ref{fig:compare3layers}(b).  The regions of strong $\pt$-breaking
are roughly associated with the resonances of the
$\mathcal{T}$-symmetric disk, as symmetry-breaking is enhanced by the
dwell time in the medium. This points to the possibility of microdisk
based amplifier-absorbers.

{\it Conclusion} -- We have shown that $\pt$-symmetric optical
scattering systems generically display spontaneous symmetry breaking,
with a unimodular phase where the S-matrix eigenstates are norm-preserving, and a broken symmetry
phase in which they are pairwise amplifying and damping with
reciprocal eigenvalue moduli.  This $\pt$-breaking transition can be
tested experimentally, using 1D heterojunction geometries (with
realistic values of the gain/loss parameter) that are distinctly
different from the paraxial geometries previously suggested for
observing $\pt$ symmetry breaking in optics
\cite{Makris,Musslimani,Guo,Ruter}.  In our analysis, we have
neglected the role of the noise due to amplified spontaneous emission, which may be 
significant at the singular CPA-laser points
\cite{Schomerus}.  However this noise should not preclude observation of
experimental signatures of CPA-lasing, e.g.~the
doubling of the free spectral range relative to the passive cavity.
Moreover one may study the inteferometric amplifying behavior in the
broken symmetry phase well below the CPA-laser points, where the noise
will be substantially smaller.

We thank S.~Longhi and A.~Cerjan for useful discussions.  This work
was partially supported by NSF Grant No. DMR-0908437, and by seed
funding from the Yale NSF-MRSEC (DMR-0520495).

\end{document}